\documentclass[aps,pra,preprint,preprintnumbers,amsmath,amssymb,showpacs]{revtex4-1}
\usepackage{graphicx}
\usepackage{color}
\usepackage{hyperref}
\usepackage{ulem}
\begin{document}

\title{Fully differential cross sections for singly ionizing 1-MeV $p$+He collisions at small momentum transfer: Beyond the first Born approximation}

\author{O.~Chuluunbaatar$^{1,2}$}
\author{S.~A.~Zaytsev$^3$}
\author{K.~A.~Kouzakov$^4$}
\author{A.~Galstyan$^5$}
\author{V.~L.~Shablov$^{6}$}
\author{Yu.~V.~Popov$^{7,1}$}
\email{popov@srd.sinp.msu.ru}

\affiliation{$^1$Joint Institute for Nuclear Research, Dubna,
Moscow region 141980, Russia}

\affiliation{$^2$Institute of Mathematics, National University of
Mongolia, UlaanBaatar, Mongolia}

\affiliation{$^3$Department of Physics, Pacific State University,
Tikhookeanskaya 136, Khabarovsk 680035, Russia}

\affiliation{$^4$Department of Nuclear Physics and Quantum Theory
of Collisions, Faculty of Physics, Lomonosov Moscow State
University, Moscow 119991, Russia}

\affiliation{$^5$Institute of Condensed Matter and Nanosciences,
Universit\'e Catholique de Louvain, 2 chemin du cyclotron, Box
L7.01.07,  B-1348 Louvain-la-Neuve, Belgium }

\affiliation{$^6$Obninsk Institute for Nuclear Power Engineering
of the National Research Nuclear University MEPhI, Russia}

\affiliation{$^7$Skobeltsyn Institute of Nuclear Physics,
Lomonosov Moscow State University, Moscow 119991, Russia}

\pacs{34.50.Fa, 52.20.Hv}

%
\begin{abstract}
We present calculations of the electron angular distributions in
the single ionization of helium by 1-MeV proton impact at momentum
transfer of 0.75 a.u. and ejected-electron energy of 6.5 eV. The
results using the first and second Born approximations and the 3C
model with different trial helium functions are compared to the
experimental data. A good agreement between theory and experiment
is found in the case of the 3C final state and a strongly
correlated helium wave function. The electron-electron
correlations in the He atom are found to influence the ratio of
the binary and recoil peak intensities.

\end{abstract}
\maketitle

%
\section{Introduction}
\label{intro}
Single ionization of atomic targets by fast ions in a perturbative
regime has been attracting much interest from experimentalists and
theorists in recent years. This is largely due to a remarkable
progress of the technique known as cold target recoil ion momentum
spectroscopy (COLTRIMS)~\cite{ulrich97,doerner00,ulrich03}.
COLTRIMS is a reaction microscope that allows one to measure in
coincidence with the residual ion three-dimensional angular
distributions of electrons emitted in the ionizing ion-atom
collisions at given values of energy and momentum transfer with
unprecedented precision. Thus, different theoretical approaches,
in particular, those based on the plane wave first Born
approximation (PWFBA), can be very robustly tested by comparing
their predictions for the fully differential cross sections (FDCS)
with the COLTRIMS data.

When comparing theory and experiment, the one of the most
interesting angular regions appears to be the one separating the
binary and recoil peaks, where a distinct node is generally
predicted by perturbation treatments. It is the region where
marked discrepancies between theory and experiment were found in
the case of singly ionizing 100-MeV/u C$^{6+}+$He collisions at
momentum transfer of 0.75\,a.u.~\cite{schulz03}. The node in the
measured electron emission pattern was much less pronounced than
that anticipated by theory. This finding was particularly
surprising given the fact that the measurements were carried out
under kinematical conditions which are believed to be perfectly
suitable for applicability of perturbative approaches: (i)
$|Z_p|/v_p=0.1$\,a.u., where $Z_p$ and $v_p$ are the projectile
charge and velocity, respectively, and (ii) small energy- and
momentum-transfer values. Further discussions involved various
attempts to explain the source of the discrepancies in the nodal
structure, ranging from
higher-order~\cite{schulz03,madison02,voitkiv03,walters10,kouzakov12}
and non-perturbative
mechanisms~\cite{ciappina06,harris07,walters10,mcgovern10,colgan11}
to experimental uncertainties~\cite{fiol06,kouzakov12} and
so-called projectile coherence effects~\cite{wang12}. Though the
explanation due to experimental uncertainties alone was refuted
in~\cite{durr07}, the very recent 1-MeV $p+$He experiment at
momentum transfer of 0.75\,a.u.~\cite{gassert16}, which has been
performed with the highest momentum resolution ever achieved in
such ionizing ion-atom collisions, has exhibited a well-pronounced
nodal structure.

It should be noted that the analysis of the data of that
experiment has not revealed clear footprints of the projectile
coherence effects. At the same time, according to the criteria
formulated in Wang et al.~\cite{wang12}, the calculated transverse
coherence lengths for a rectangularly collimated projectile beam
in the discussed experiment were in the intermediate regime, i.e.
not large enough to yield a coherent beam and not sufficiently
small to yield an incoherent beam. This means that the problem of
the projectile coherence can still be claimed as not completely
resolved (see, for instance, the recent
work~\cite{Arthanayaka16}). The most recent theoretical analyses
and discussions devoted to this problem one can find in
Refs.~\cite{kouzakov17,navarrete17,nagy17} and references therein.

For shedding light onto the origin of the disagreement between
theory and experiment one must also analyse the role of
theoretical uncertainties, which are due to inaccuracy of
theoretical models employed for calculations of FDCS. These
uncertainties arise from approximations involved in the treatment
both of the collision mechanism and of the initial and final
states of the target. Therefore, apart from delivering details of
experiment that determine momentum resolution and
projectile-coherence effects, the purpose of this work is to
present theoretical calculations of FDCS using different models of
the initial and final helium states as well as different models of
the collision mechanism. Currently there are two main theoretical
approaches to ionization of a quantum target by impact of a proton
(positive ion). These are the continuum distorted wave-eikonal
initial state (CDW-EIS) method in the semiclassical and fully
quantum formulations respectively. In the semiclassical approach,
one treats the proton as a classical particle moving along a
straight-line trajectory with an impact parameter $\rho$ and
velocity $v_p$, which induces time-dependent perturbation of the
quantum target. The time-dependent Schr\"odinger equation for the
target is solved with the appropriate boundary condition at $t\to
-\infty$. In the quantum approach, one solves a stationary quantum
scattering problem for all particles involved in the ionization
reaction. An example of semiclassical calculations for the
kinematical domain discussed in the present work can be found in
\cite{voitkiv17}.

To the best of our knowledge, for the first time the stationary
formulation of the CDW-EIS method was given in
Refs.~\cite{dodd66,gayet72}. It was further developed in later
works (see, for instance, Refs.~\cite{gasaneo96,ciappina06}).
Though some of them dealt with charge-transfer reactions, the
quantum CDW-EIS method was elaborated in these works rather fully.
The semiclassical CDW-EIS method was formulated for the first
time, to the best of our knowledge, in Ref.~\cite{cheshire64}
(strictly speaking, J. Cheshire formulated a CDW-CDW model). Later
it was developed in Refs.~\cite{crothers82,crothers83}, which are
mainly cited by the authors of more recent works (see, for
instance, those by the group of Rivarola {\it et
al.}~\cite{ciappina14,monti13}). Both formulations, along with
their relation to each other within a rigorous mathematical
approach, are discussed in the well-known review
articles~\cite{belkic79,crothers93}.

Besides the CDW-EIS method there are more traditional approaches
in the literature, where in the exact post-form matrix element for
the transition amplitude the full final-state wave function is
replaced with its various approximations. It is believed that the
closer the asymptotics of the approximate wave functions to the
exact asymptotics at large distances between the particles the
better such approximations work. In this regard, it is worth to
mention the works of Madison~{\it et al.} (see, for instance,
Refs.~\cite{madison02,harris07}). Clearly, all the above-mentioned
methods and approaches are related to each other in one way or
another, and they well explain discrepancies between PWFBA and
experiment at high incident energy of a projectile (proton). It
should be noted that very often they yield practically the same
results in the most of kinematical regions of ionization reactions
with MeV's protons.

The special case of the quantum CDW-EIS is the so-called 3C (or
BBK) model~\cite{BBK}. In this model, three Coulomb continuum
functions are employed for description of the final-state
interactions. It should be noted that most of the matrix-element
calculations are performed in real space using trial wave
functions of the target that are rather simple. The latter is due
to the complexity of the calculations (see, for instance,
Ref.~\cite{Mondal16}).

In this work, we present calculations of FDCS beyond the PWFBA
theory in order to explain the shift of the measured binary and
recoil peaks in the scattering plane by few degrees towards the
incident proton direction with respect to the PWFBA
predictions~\cite{gassert16}. For this purpose we examine the
plane wave second Born approximation (PWSBA) and the well-known 3C
(or BBK) model~\cite{BBK}. The calculations are performed in
momentum space, what allows one to inspect different models of the
helium ground state, including the strongly correlated ones. Three
different ground-state wave functions of He are used in the
calculations: (i) the loosely correlated Roothaan-Hartree-Fock
function (RHF)~\cite{RHF}, (ii) the trial wave function of
Silverman-Platas-Matsen (SPM)~\cite{SPM} from the configuration
interaction family, and (iii) a strongly correlated function
(CF)~\cite{CF} of the Bonham-Kohl type. For accurate comparisons
with experiment, all theoretical values are convoluted with
experimental uncertainties.

The paper is organized as follows. In Sec.~\ref{theo}, we
formulate different theoretical models and approximations for the
considered process. Then, in Sec.~\ref{res}, we compare
experimental and theoretical results. The conclusions are drawn in
Sec.~\ref{concl}. Atomic units (a.u.), in which $\hbar=e=m_e=1$, are
used throughout unless otherwise specified.

\section{Theory}
\label{theo}
In Ref.~\cite{gassert16}, we calculated FDCS for the discussed
ionization reaction using the PWFBA. We designate the incident and
final proton momenta by $\vec p_i$ and $\vec p_s$, respectively,
the electron momentum by $\vec k_e$, the final ion momentum by
$\vec K_{ion}$. In the experiment, the momentum transfer $Q$ is
relatively small, $Q=0.75$ a.u., and the ejected-electron is
$E_e=6.5$ eV (0.24 a.u.). The momentum conservation law
\begin{equation}
\vec Q=\vec p_i-\vec p_s=\vec k_e+\vec K_{ion} \label{(1)}
\end{equation}
shows that the velocity of the residual ion $K_{ion}/(m_N+1)$ is
practically negligible (the mass of the He atom is $m_N\approx
4m_p=7344.6$ a.u.), what allows us to choose the stationary He
nucleus as a center of the laboratory coordinate system.

From the energy conservation law
\begin{equation}
E=\frac{p^2_i}{2m_p}+\varepsilon_0^{He}=\frac{(\vec p_i-\vec
Q)^2}{2m_p}+\varepsilon_0^{He^+}+\frac{k^2_e}{2}+\frac{K^2_{ion}}{2(m_N+1)}
\label{(2)}
\end{equation}
one obtains the $z$-component of the momentum transfer as
$Q_z=(-\varepsilon_0^{He}+\varepsilon_0^{He^+}+E_e)/v_p=0.18$ a.u.
(the $z$ axis is directed along the initial proton momentum
$\vec{p}_i$). The transverse component is $Q_\perp\approx m_p
v_p\theta_s=0.73$ a.u. ($\theta_s$ is the scattering angle of the
proton), with $v_p=p_i/m_p$ being the proton velocity. The He and
He$^+$ kinetic energies, $Q^2/2m_p$ and $K^2_{ion}/2(m_N+1)$, are
neglected in Eq.~(\ref{(2)}).

The general expression for FDCS reads
\begin{equation}
\frac{d^3\sigma}{dE_ed\Omega_ed\Omega_s}=
k_e\frac{m^2_p}{(2\pi)^5}|T_{fi}|^2. \label{(3)}
\end{equation}
This form is different from that in Ref.~\cite{madison02}. It is
correct for small momentum transfers, when the velocity of the
recoil ion is practically zero. In this case we can place the
center of coordinates at the ion and use the proton mass $m_p$
instead of the reduced mass $\mu_{pN}=m_pm_N/(m_p+m_N)$~\cite{LL}.

The final state of the considered reaction
contains three charged fragments: $p$, $e$ and the He$^+$ ion.
In general, the Dollard asymptotic conditions~\cite{Dollard} should
be taken into account. Within the PWFBA these conditions are
discarded. Below we consider a correlated 3C final-state wave
function $\Phi_f$~\cite{BBK} which satisfies these conditions.

As indicated in the introduction section, three initial trial
helium wave functions $\Phi_0^{He}$ are employed: a weakly
correlated Roothaan-Hartree-Fock (RHF) \cite{RHF} function,
$\varepsilon_0^{RHF}=-2.8617$\,a.u., a simple
Silverman-Platas-Matsen (SPM) function~\cite{SPM} of the
configuration interaction family,
$\varepsilon_0^{SPM}=-2.8952$\,a.u., and a strongly correlated
function (CF)~\cite{CF} which explicitly depends on the $r_{12}$
distance between electrons in helium,
$\varepsilon_0^{CF}=-2.903724$\,a.u.. The helium energy of last
function is very close to the experimental value,
$\varepsilon_0^{exp}=-2.903724$\,a.u.

\subsection*{General formulas for the 3C wave function}
In this approximation, the matrix element has the following form (the electron
identity is taken into account by the factor of $\sqrt{2}$):
\begin{eqnarray}
T_{fi}^{\rm 3C}&=&\sqrt{2}\int d^3R d^3r_1 d^3 r_2\  e^{i\vec
R\vec p_i} \Psi_f^{(-*)}(\vec R,\vec r_1,\vec r_2; \vec p_s, \vec
k_e){\Phi}_0^{He}(\vec r_1,\vec r_2)
\nonumber\\
&{}&\times
\left[\frac{2}{R}-\frac{1}{|\vec R-\vec r_1|}-\frac{1}{|\vec
R-\vec r_2|}\right], \label{(4)}
\end{eqnarray}
where the final-state wave function is
\begin{eqnarray}
\Psi_f^{-*}(\vec R,\vec r_1,\vec r_2)&=&e^{-i\vec R\vec
p_s}\varphi_0^{He^+}(\vec r_2){\tilde\phi}^{-*}(\vec k_e,\vec
r_1;-1)\, \exp\left(\frac{\pi}{2|\vec v_p-\vec
k_e|}\right)\Gamma\left(1-i\frac{1}{|\vec v_p-\vec k_e|}\right)\nonumber\\
&{}&\times
{_1}F_1\left[i\frac{1}{|\vec v_p-\vec k_e|},1; i(|\vec R-\vec
r_1||\vec v_p-\vec k_e|+(\vec R-\vec r_1)\cdot(\vec v_p-\vec
k_e))\right]\nonumber\\
&{}&\times\exp\left(-\frac{\pi}{2v_p}\right)\Gamma\left(1+i\frac{1}{v_p}\right)
{_1}F_1\left[-i\frac{1}{v_p},1; i(Rp_{r}+\vec R\cdot\vec
p_{r})\right]. \label{(5)}
\end{eqnarray}
This is the 3C (BBK) function. Here $\vec r_1$, $\vec r_2$, and
$\vec R$ are the electron and proton positions with respect to the
nucleus,
$$
\vec p_r=[(4m_p+1)\vec p_s-m_p\vec K_{ion}]/(5m_p+1)\approx
(4m_p/5)\vec v_p+(1/5)\vec k_e-\vec Q,
$$
and $m_p=1836.15$\,a.u. is the proton mass. Further, $\vec
p_i=m_p\vec v_p$, $v_p=6.35$\,a.u.,
$$
{\tilde\phi}^{-*}(\vec q,\vec r;Z)=e^{-\pi\xi/2}\Gamma(1+i\xi)
e^{-i\vec q\cdot\vec r} {_1}F_1(-i\xi,1;iqr+i{\vec q}\cdot{\vec
r})
$$
is the Coulomb continuum function, with $\xi=Z/v_q$, where
$Z=Z_jZ_k$ ($Z_j$ and $Z_k$ are charges of interacting particles).
Finally, the ion ground state is
\begin{equation}
\varphi_0^{He^+}(\vec r_2)=\sqrt{\frac{8}{\pi}}\exp(-2r_2).
\label{(6)}
\end{equation}
The PWFBA follows from Eq.~(\ref{(5)}) if one formally sets
$1/v_p=0$ and $1/|\vec v_p-\vec k_e|=0$.

Let us explain our choice of the charges in the 3C wave function,
especially in the Coulomb function ${\tilde\phi}^{-*}(\vec
k_e,\vec r_1;-1)$, which describes an escape of the electron from
the helium atom. The 3C function is basically asymptotic,
providing the correct Dollard asymptotic behavior. If we would
like to treat the motion of the escaping electron in the realistic
potential, which is much more complex than a Coulomb potential,
then we must deal with a full four-body problem inside the atomic
target. This would bring about great difficulties from the
viewpoint of numerical computations.

For numerical calculations on the basis of~(\ref{(4)}) it is
convenient to perform its Fourier transformation that reduces
significantly the number of integrations.  The amplitude  splits
into the sum of three integrals, $T_{fi}=A_1+A_2+A_3$. We treat
them separately and obtain
\begin{eqnarray}
 A_1&=&2\sqrt2\int\frac{d^3 p}{(2\pi)^3}\
\phi^{-*}(\vec v_p-\vec k_e,\vec p+\vec v_p-\vec k_e;-1)I(\vec
p_r,\vec Q+\vec p_r-\vec p; 1)G(\vec k_e,\vec p,0), \label{A1}\\
A_2&=& -\sqrt2\int\frac{d^3 p}{(2\pi)^3}\ I(\vec v_p-\vec k_e,\vec
p+\vec v_p-\vec k_e;-1)\phi^{-*}(\vec p_r,\vec Q+\vec p_r-\vec
p;1)G(\vec k_e,\vec p,0), \label{A2}\\
A_3&=& -4\pi\sqrt2\int\frac{d^3 p_1}{(2\pi)^3}\frac{d^3
p_2}{(2\pi)^3
p_2^2}\ \phi^{-*}(\vec v_p-\vec k_e,\vec p_1+\vec v_p-\vec k_e;-1)\nonumber\\
&{}&\times\ \phi^{-*}(\vec p_r,\vec Q+\vec p_r-\vec p_1-\vec
p_2;1)G(\vec k_e,\vec p_1,\vec p_2).\label{A3}
\end{eqnarray}
In (\ref{A1}) $\div$ (\ref{A3}) we have used the following
notations:
\begin{equation}
I(\vec q,\vec p;
Z)=\lim_{\lambda\to+0}\int\frac{d^3r}{r}e^{-\lambda
r}{\tilde\phi}^{-*}(\vec q, \vec r; Z) e^{i\vec p\cdot\vec
r}=\lim_{\lambda\to+0}\ 4\pi\
e^{-\pi\xi/2}\Gamma(1+i\xi)\frac{[p^2-(q+i\lambda)^2]^{i\xi}}{[(\vec
p-\vec q)^2+\lambda^2]^{(1+i\xi)}}, \label{I}
\end{equation}
$$
\phi^{-*}(\vec p,\vec q;Z)=-\frac{\partial}{\partial\lambda}\
I(\vec p,\vec q; Z),
$$
and
\begin{equation}
G(\vec k,\vec q_1,\vec q_2)=\int d^3r_1 d^3 r_2\
{\tilde\phi}^{-*}(\vec k,\vec r_1;-1)e^{i\vec q_1\cdot\vec r_1}
\varphi_0^{He^+}(\vec r_2)e^{i\vec q_2\cdot\vec
r_2}{\Phi}_0^{He}(\vec r_1,\vec r_2). \label{G}
\end{equation}
For most of the He ground-state models the function $G(\vec
k,\vec q_1,\vec q_2)$ is the analytical function or an integral of
much lower dimension.

Some comments should be made about details of calculations. We
investigated the behavior of the results of integration of
(\ref{A1}) $\div$ (\ref{A3}) with a finite parameter $\lambda$. It
was found that at $\lambda= 10^{-2}\div 10^{-4}$ we obtain very
good stability of calculations and convergence. Results with
$\lambda=10^{-3}$ and $\lambda=10^{-4}$ are practically
indistinguishable. Finally, the 3D, 4D and 7D integrals (depending
on the trial helium functions) are calculated numerically using
the Fortran code Cuhre \cite{cuba}.

\subsection*{PWSBA and the closure approximation}
We use the abbreviation PWSBA to emphasize that the proton is
described by the plane wave. It follows from~(\ref{(4)}) if in the
expansion of the final-state wave function in powers of $1/v_p$
and $1/|\vec{v}_p-\vec{k}_e|$ we retain only the zeroth- and
first-order terms:
\begin{equation}
{\cal T}(\vec v_p,\vec Q,\vec k_e)= T^{PWFBA}(\vec Q,\vec k_e)+
T^{PWSBA}(\vec v_p,\vec Q,\vec k_e). \label{(12)}
\end{equation}
The PWSBA term is given by
\begin{eqnarray}
T^{\rm PWSBA}_{fi}&=&\frac{2\sqrt 2}{\pi}\sum_\alpha\int
\frac{d^3x}{x^2(\vec Q-\vec x)^2[\vec v_p\cdot\vec
x+\varepsilon_0^{He}-\varepsilon^{He}_\alpha+i0]}
\nonumber\\
&{}&\times\int d^3r_1 d^3 r_2 \Phi_f^{He(-*)}(\vec r_1,\vec r_2;
\vec k_e)\left[2-e^{i(\vec Q-\vec x)\vec{r}_1}-e^{i(\vec Q-\vec
x)\vec {r}_2}\right]\Phi_{\alpha}^{He(-)}(\vec r_1,\vec r_2)
\nonumber\\
&{}&\times\int d^3r_1' d^3 r_2'
\Phi_{\alpha}^{He(-*)}(\vec{r}_1',\vec r_2') \left[2-e^{i\vec
x\vec{r}_1'}-e^{i\vec x\vec {r}_2'}\right]\Phi_0^{He}(\vec
r_1',\vec r_2'). \label{(13)}
\end{eqnarray}
The sum in Eq.~(\ref{(13)}) runs over all helium eigenstates including single and double continuum.

Since the proton velocity is large, we use the closure
approximation $\varepsilon^{He}_\alpha-\varepsilon_0^{He} \to
{\bar E}>0$. In this case we get
\begin{eqnarray}
T^{\rm PWSBA}_{fi}&=&\frac{2\sqrt 2}{\pi}\int \frac{d^3x}{x^2(\vec
Q-\vec x)^2[\vec v_0\cdot\vec x-{\bar E}+i0]}
\nonumber\\
&{}&\times\int d^3r_1 d^3 r_2 \Phi_f^{He(-*)}(\vec r_1,\vec r_2;
\vec k_e)[2-e^{i(\vec Q-\vec x)\vec{r}_1}-e^{i(\vec Q-\vec x)\vec
{r}_2}][2-e^{i\vec x\vec{r}_1}-e^{i\vec x\vec
{r}_2}]\Phi_0^{He}(\vec r_1,\vec r_2).\nonumber\\
 \label{(14)}
\end{eqnarray}
For the initial state we use the Hylleraas wave function (the
ground-state energy is $\varepsilon_0^{Hy}=-2.8477$ a.u.)
\begin{equation}
\label{Hy} {\Phi}_0^{He}(\vec r_1,\vec r_2)=\phi(r_1)\phi(r_2),
\qquad \phi(r)=\sqrt{\frac{Z_{h}^3}{\pi}}e^{-Z_{h}r}, \qquad
Z_{h}=27/16.
\end{equation}
The final single-electron state is taken in the simplest form
$$
\Phi_f^{He(-*)}(\vec r_1,\vec r_2; \vec
k_e)={\tilde\phi}^{-*}(\vec k_e,\vec r_1;-1)\varphi^{He^+}(r_2).
$$

Our formulation of PWSBA is very close to that in the
work~\cite{walters10}. It is interesting to note that numerically
and theoretically the PWSBA results using the closure
approximation are very similar to those obtained within the
eikonal wave Born approximation (EWBA). Earlier this particular
approach was formulated in Ref.~\cite{EWBA}. It was noted that all
the interactions involved in the perturbation contribute to the
corresponding matrix element on an equal footing, that is, not
only the projectile-nucleus interaction appears to be important.

\section{Results and discussion}
\label{res}
In this section we present results of numerical calculations for
FDCS as a function of the electron scattering angle $\theta_e$ in
the coplanar geometry, which were obtained using the theoretical
approaches outlined in the previous section. For comparison with
experiment, the theoretical results were convoluted with an
angular resolution of 5$^\circ$, both in polar and in azimuthal
angles, and averaged over the out-of-plane electron angle
$\phi_e=0\pm10^\circ$ (see details in the Appendix~\ref{app}). As
the experiment was performed on an arbitrary intensity scale, the
experimental values were normalized to the convoluted and averaged
FDCS in the case of the 3C model~(\ref{(5)}) and the CF
ground-state function of He.

\begin{figure}
\includegraphics[scale=0.6]{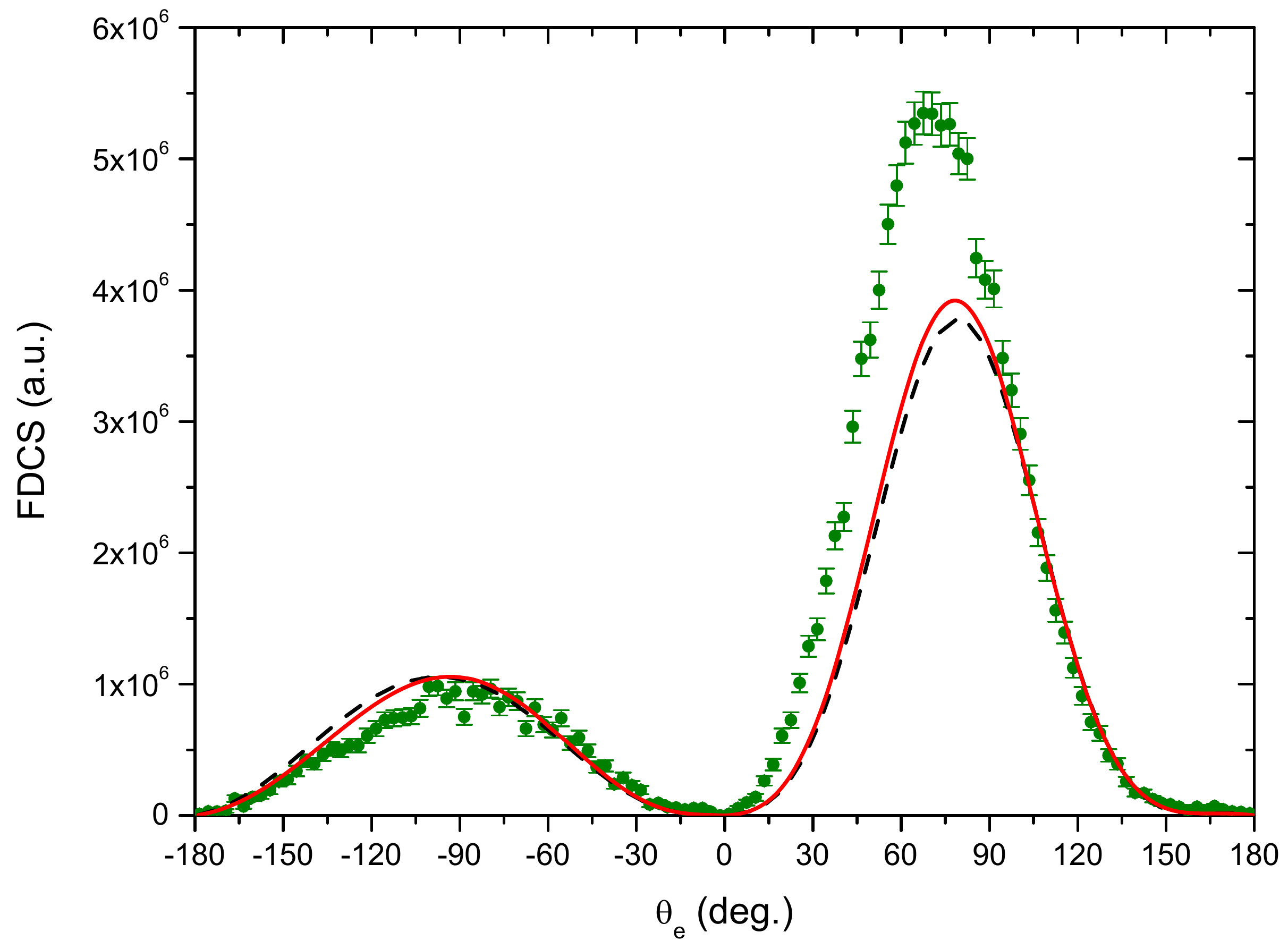}
\caption{\label{figSBA} (Color online) FDCS in coplanar geometry
using the PWFBA (black dashed line) and PWSBA (red solid line)
with the Hy ground state of the He atom~(\ref{Hy}). Experimental
values from \cite{gassert16} are represented by points, $\bar
E=0.9$.}
\end{figure}
\begin{figure}
\includegraphics[scale=0.6]{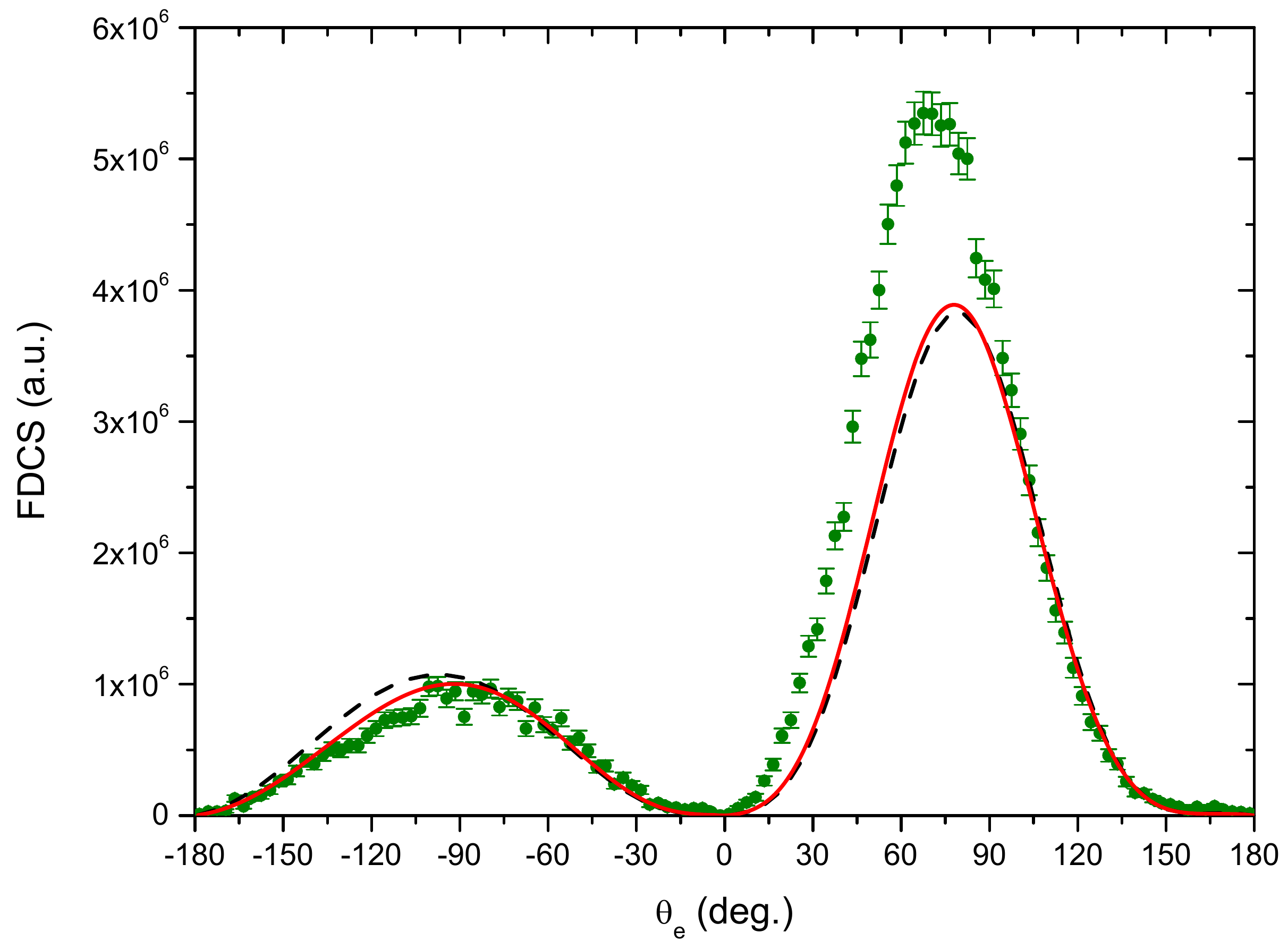}
\caption{\label{figRHF} (Color online) FDCS in coplanar geometry
using the PWFBA (black dashed line) and 3C model (red solid line)
with the RHF ground state of the He atom. Experimental values from \cite{gassert16} are
represented by points, $\lambda=10^{-3}$.}
\end{figure}

The electron angular distribution within PWFBA typically exhibits
two distinct peaks of larger and smaller intensity respectively:
the binary peak in the direction of momentum transfer and the
recoil peak in the opposite direction. In our previous
calculations~\cite{gassert16}, it was found that the measured
binary and recoil peaks in coplanar geometry are shifted by almost
10$^\circ$ towards the incident proton direction with respect to
the PWFBA theory. This is an unambiguous signature of the
collision mechanisms beyond PWFBA. In Fig.~\ref{figSBA}, the PWFBA
results using the Hy function~(\ref{Hy}) are presented along with
the PWSBA calculations and experimental values. The PWFBA and
PWSBA cross sections appear to be close to each other. At the same
time, the binary and recoil peaks in the PWSBA case are shifted
(by $\sim$3$^\circ$-5$^\circ$) relative to the PWFBA ones towards
the $\theta_e=0^\circ$ direction. Displacement of the binary peak
practically is not changed within wide domain $0.1\lesssim{\bar
E}\lesssim 1.7$, and is too small for explaining positions of the
peaks' maximums in the experimental angular distribution,
indicating that the higher-order effects, beyond PWSBA, should be
taken into account.

\begin{figure}
\includegraphics[scale=0.6]{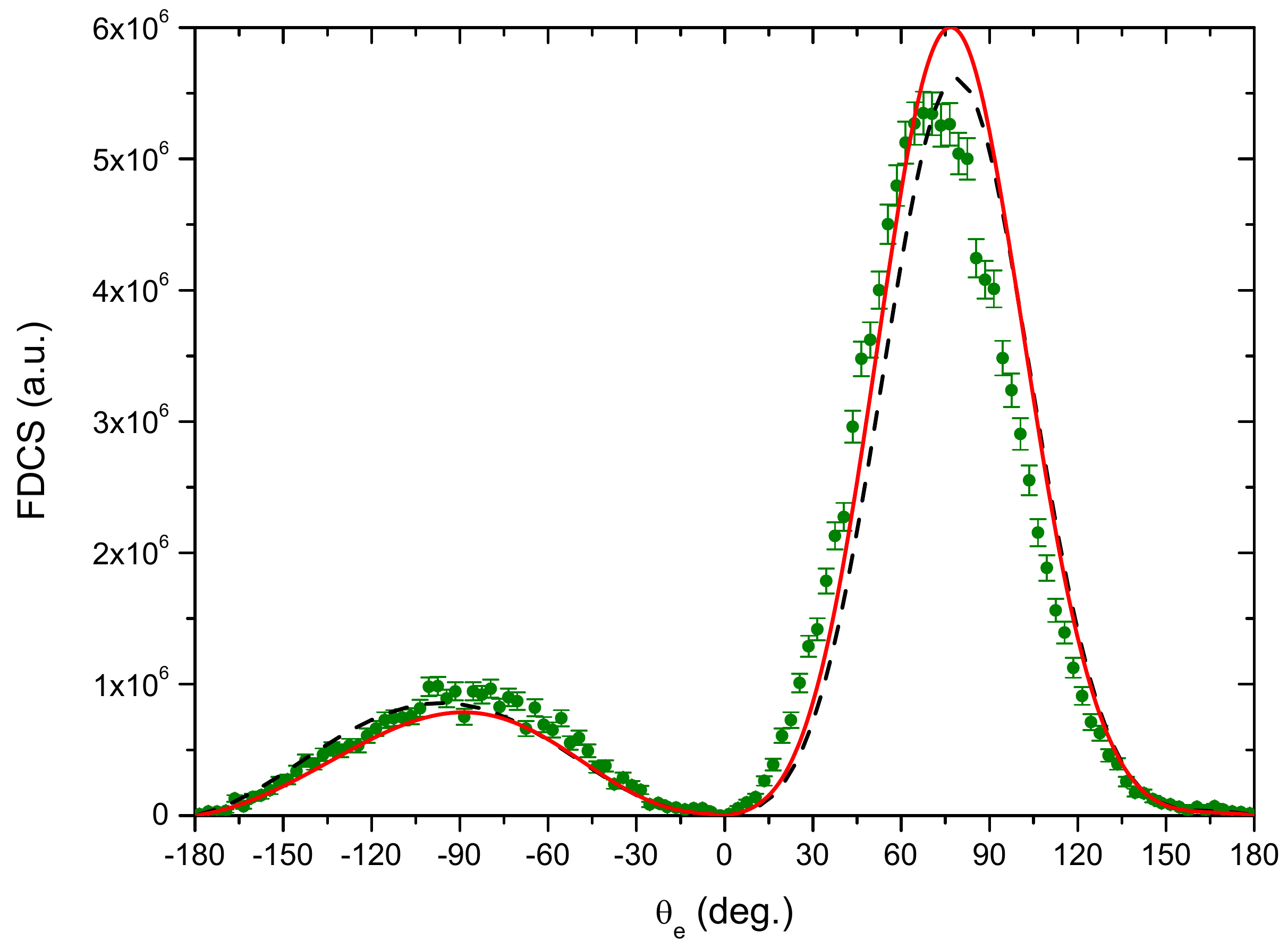}
\caption{\label{figSPM} (Color online) The same as in Fig.~\ref{figRHF},
but in the case of the SPM ground state of the He atom.}
\end{figure}

As remarked in the previous section, PWFBA and PWSBA follow from
the 3C model upon expansion in powers of $1/v_p$ and
$1/|\vec{v}_p-\vec{k}_e|$. In other words, the 3C model
effectively includes higher-order collision mechanisms unaccounted
by these approximations. Figure~\ref{figRHF} shows the PWFBA and
3C results in the case of the RHF function in comparison with
experiment. The difference between the PWFBA and 3C results is not
very significant. Similar to PWSBA, the 3C model provides the
shift of both PWFBA peaks towards the experiment. The shift for
the binary peak is much smaller than for the recoil peak and is
not enough to explain the experiment. It is clear that the ratio
of measured peak intensities (recoil/binary) does not depend on
the employed normalization of experiment. In this respect, both
the PWFBA and the 3C results substantially disagree with
experiment. This disagreement is due to a poor account for
electron-electron correlations in the He ground state with the RHF
model. The effect of electron-electron correlations in He is known
to be strong and cannot be appropriately treated by mean-field
approaches such as RHF. Even the simple SPM function of
the configuration interaction family describes this effect much
better. As can be seen from Fig.~\ref{figSPM}, it gives the value
of the recoil/binary ratio which is notably closer to the
experiment than in the RHF case. The differences between the PWFBA
and 3C results using the SPM function are similar to those in
Fig.~\ref{figRHF}.

\begin{figure}
\includegraphics[scale=0.6]{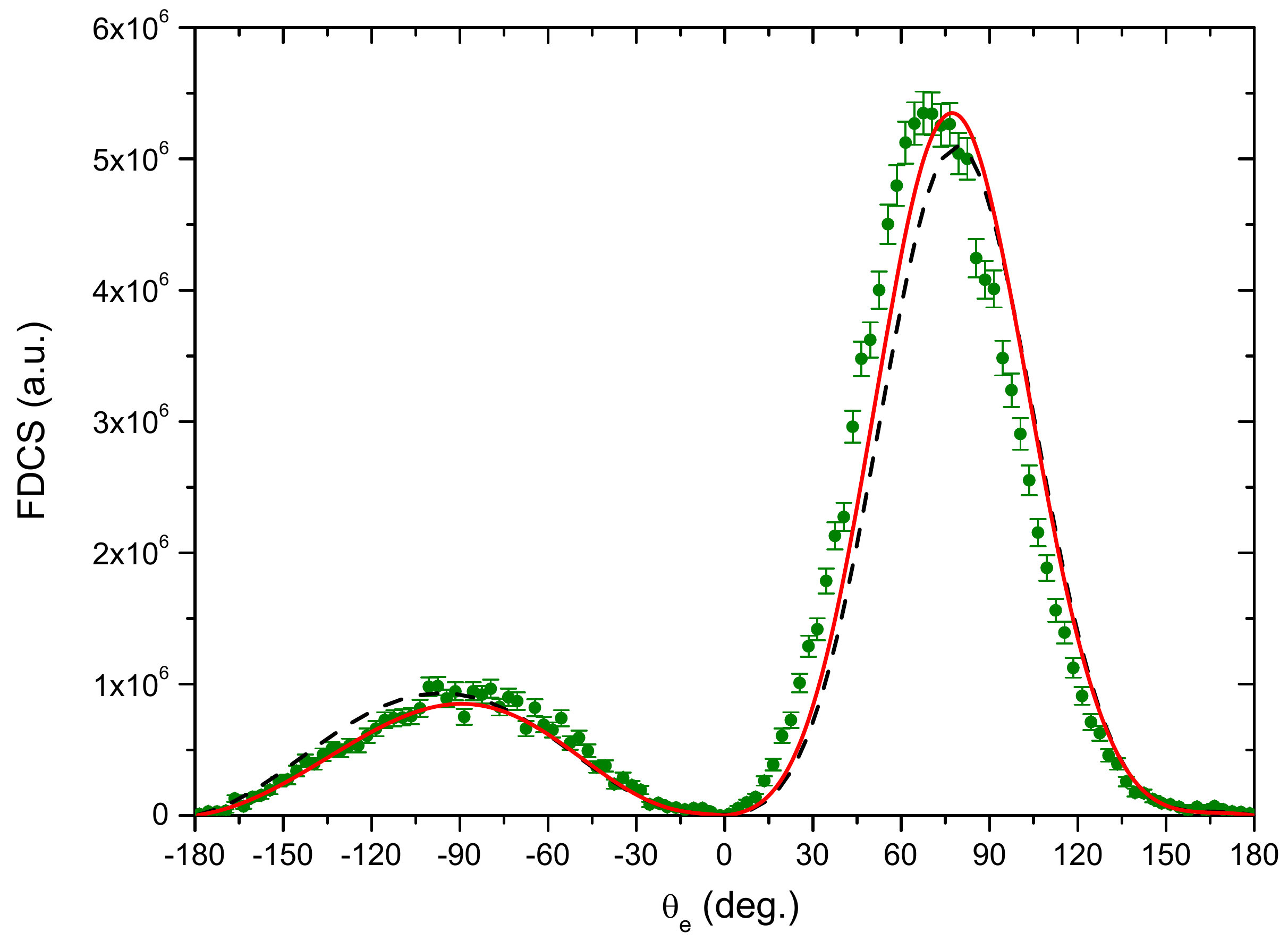}
\caption{\label{figCF} (Color online) The same as in
Fig.~\ref{figRHF}, but in the case of the CF ground state of the
He atom.}
\end{figure}

Figure~\ref{figCF} compares with experiment the PWFBA and 3C
calculations using the CF function. The 3C model gives good
agreement with experiment in terms of the recoil/binary ratio and
differs from the PWFBA results in a similar way as in
Figs.~\ref{figRHF} and~\ref{figSPM}.

We compared our in-plane results without convolution with those
reported recently in Ref.~\cite{voitkiv17} (see Fig.~1 therein).
For making the comparison, we multiplied the scale employed in
Ref.~\cite{voitkiv17} by the factor $k_ep^2_i$. Our 3C results
using the CF function fully agree with those in
Ref.~\cite{voitkiv17}. This finding is surprising because we would
expect an agreement with our results using the RHF function rather
than the CF function. Indeed, the author of Ref.~\cite{voitkiv17}
calculated the helium ground-state function using a simple
one-electron radial potential and neglecting the correlation
between electrons in helium. At the same time, the shift of the
binary peak in Ref.~\cite{voitkiv17} is the same as in the present
paper, i.e. it is not sufficient to explain the experimental
binary-peak position.

\begin{figure}
\includegraphics[scale=0.6]{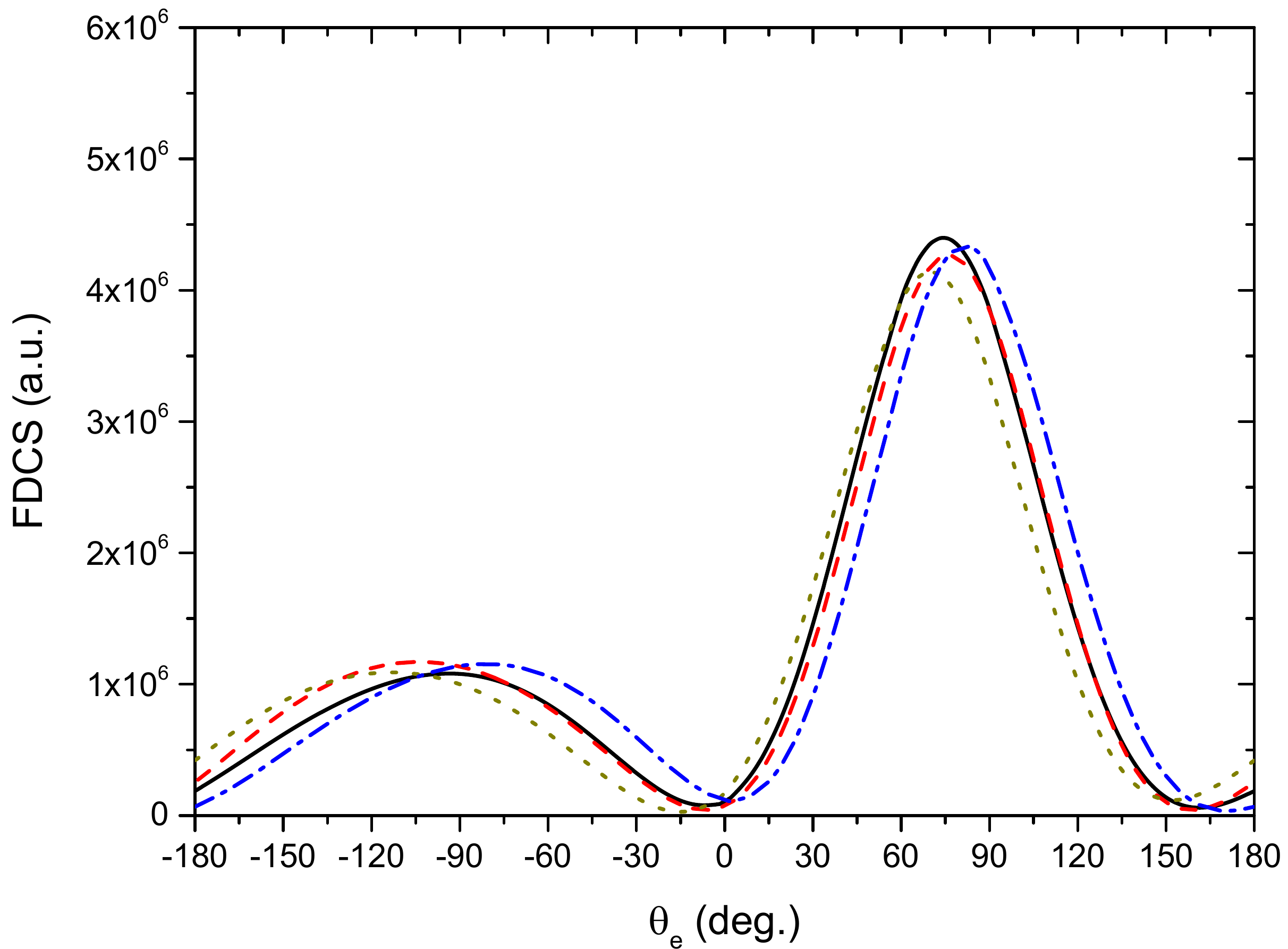}
\caption{\label{fig_test} (Color online) FDCS in the case of the
CF ground state of the He atom within PWFBA (red dashed line), 3C
model (black solid line), and the 3C model with $1/v_p=0$ (blue
dash-dotted line) and $1/|\vec{v}_p-\vec{k}_e|=0$ (bronze yellow
dotted line). $\lambda=10^{-3}$.}
\end{figure}

In order to inspect which final-state interaction is mainly
responsible for the shift of the binary and recoil peaks, we
performed the 3C calculations with the CF function in the
following situations: (i) $1/v_p=0$, i.e. no proton-ion
interaction in the final state, and (ii)
$1/|\vec{v}_p-\vec{k}_e|=0$, i.e. no proton-electron interaction
in the final state. The results of such test calculations are
shown in Fig.~\ref{fig_test}. One would expect that the shift of
the peaks is due to the proton-electron rather than the proton-ion
interaction, since the proton, after knocking-out the electron
from the atom, attracts the electron and thus should distort the
outgoing electron trajectory in the forward direction. The results
presented in Fig.~\ref{fig_test} agree with this expectation
concerning the recoil peak but contradict it regarding the binary
peak. It turns out that it is the proton-ion interaction that is
responsible for the shift of the binary peak towards smaller
$\theta_e$ values, while the proton-electron interaction shifts
the binary peak in the opposite direction. Moreover, the shift in
the $1/v_p=0$ case is large enough to explain the position of the
binary peak in experiment. It is worth mentioning that the
conclusion that for small momentum transfers it is the
proton-nucleus interaction that mostly leads to a shift of the
binary peak has been formulated earlier in Ref.~\cite{Rolla}.

\section{Summary and conclusions}
\label{concl}
Theoretical calculations beyond the PWFBA theory have been
presented and compared with experiment in coplanar geometry. The
roles of higher-order collision mechanisms have been examined
using the PWSBA and 3C approaches. Three different ground-state
wave functions of He have been employed in the calculations. It
has been shown that for explaining the recoil peak/binary peak
ratio it is necessary to account for the strong effect of
electron-electron correlations in the ground state of the He atom.
The PWSBA treatment has been found to be insufficient for
describing the shift of the binary and recoil peaks in the
electron angular distribution with respect to the PWFBA
prediction. In this respect, the 3C model reasonably agrees with
experiment in the recoil peak, but a discrepancy of few degrees with
experiment still remains in the case of the binary peak. Our test
calculations within the 3C model have yielded a counter-intuitive result, namely
that the shift of the binary peak towards smaller electron angles
is due to the proton-ion interaction in the final state. This
finding contradicts a naive expectation that this shift is due to
the proton-electron interaction in the final state. Surprisingly,
the latter interaction shifts the binary peak in the opposite
direction. Thus, further theoretical studies are needed to explain
the discrepancy between theory and experiment in coplanar
geometry.

\begin{acknowledgments}
The authors are grateful to Markus Sch\"offler and Reinhard
D\"orner for useful discussions and help. The present research
benefited from computational resources of the Central Information
and Computer Complex and the HybriLIT heterogeneous computing
cluster of the Joint Institute for Nuclear Research, as well as
from supercomputer Lomonosov of Moscow State University. Yu.~P. is
grateful to Russian Foundation for Basic Research (RFBR) for the
financial support under the grant No.~16-02-00049-a. O.~Ch.
acknowledges support from the Hulubei-Meshcheryakov program
JINR-Romania. A.~G. is "aspirant au Fonds de la Recherche
Scientifique (F.R.S-FNRS)"
\end{acknowledgments}

\appendix

\section{Convolution of theoretical values with experimental uncertainties}
\label{app}

Let the 2D cross section $\sigma^{2D}(\theta,\phi)$ to be
tabulated in the points $(\theta_i,\phi_j)$,
$i=1,\ldots,M,\,j=1,\ldots,N$. In the convolution procedure, we
use the following Gaussian function at each point $(\theta_i,\phi_j)$:
\begin{eqnarray}
G^{2D}(\theta,\phi,\theta_i,\phi_j)=\frac{1}{w_\theta\sqrt{2\pi}}\frac{1}{w_\phi\sqrt{2\pi}}
\,\exp\left(-\frac{1}{2}\left(\frac{\theta-\theta_i}{w_\theta}\right)^2-\frac{1}{2}
\left(\frac{\phi-\phi_j}{w_\phi}\right)^2\right)
\end{eqnarray}
with full width at half maximum (FWHM)
\begin{eqnarray}
{\rm FWHM}=2\sqrt{2\ln(2)}\,w_\theta=2\sqrt{2\ln(2)}\,w_\phi. 
\end{eqnarray}
The convoluted discrete FDCS, $\sigma^{2D}_C(\theta,\phi)$, is
constructed at the given point $(\theta_i,\phi_j)$ according to
\begin{eqnarray}
\sigma^{2D}_C(\theta_{i},\phi_{j})=\sum_{k=\max(-K,1-i)}^{\min(K,M-i)}\sum_{l=
\max(-L,1-j)}^{\min(L,N-j)} G^{2D}(\theta_{i},\phi_{j},\theta_{i+k},\phi_{j+l})
\sigma^{2D}(\theta_{i+k},\phi_{j+l}).
\end{eqnarray}
Here
\begin{eqnarray}
K=[{\rm FWHM}/h_\theta]+1, \qquad L=[{\rm FWHM}/h_\phi]+1,
\end{eqnarray}
where $h_\theta$ and $h_\phi$ are steps of the $\theta$ and $\phi$
angles, and $[x]$ denotes an integer part of $x$.

In the present case, we used $M=361$, $N=21$,
$h_\theta=h_\phi=1^\circ$, $\theta_i=(i-1)h_\theta$,
$\phi_j=-10^\circ+(i-1)h_\phi$ and FWHM$=5$. The convoluted FDCS
are further averaged over angular windows, so that one has for
coplanar geometry
\begin{eqnarray}
\sigma^{1D}_C(\theta,\phi=0^\circ)=|\sin(\theta)|\int^{10^\circ}_{-10^\circ}d\phi'\sigma^{2D}_C(\theta,\phi').
\end{eqnarray}
The integral in (A5) is calculated numerically using points
$\phi_j$, $j=1,\ldots,N$ for given $\theta=\theta_i$,
$i=1,\ldots,M$.

\end{document}